# Predicting and Explaining Customer Data Sharing in the Open Banking

João B. G. de Brito[1][0000-0002-4645-5217], Rodrigo Heldt[1][0000-0002-3435-8962], Cleo S. Silveira[1][0000-0001-5340-9107], Matthias Bogaert[2][0000-0002-4502-0764], Guilherme B. Bucco[1][0000-0002-3813-8406], Fernando B. Luce[1][0000-0003-2589-2845], João L. Becker[3][0000-0003-4176-7374], Filipe J. Zabala[1][0000-0002-5501-0877], and Michel J. Anzanello[1][0000-0002-4421-7004]

[1] Federal University of Rio Grande do Sul, Porto Alegre, Brazil
[2] Ghent University, Ghent, Belgium
[3] Fundação Getúlio Vargas, São Paulo, Brazil

joaobatista.goncalvesdebrito@ugent.be

**Abstract.** The emergence of Open Banking represents a significant shift in financial data management, influencing financial institutions' market dynamics and marketing strategies. This increased competition creates opportunities and challenges, as institutions manage data inflow to improve products and services while mitigating data outflow that could aid competitors. This study introduces a framework to predict customers' propensity to share data via Open Banking and interprets this behavior through Explanatory Model Analysis (EMA). Using data from a large Brazilian financial institution with approximately 3.2 million customers, a hybrid data balancing strategy incorporating ADASYN and NEARMISS techniques was employed to address the infrequency of data sharing and enhance the training of XGBoost models. These models accurately predicted customer data sharing, achieving 91.39% accuracy for inflow and 91.53% for outflow. The EMA phase combined the Shapley Additive Explanations (SHAP) method with the Classification and Regression Tree (CART) technique, revealing the most influential features on customer decisions. Key features included the number of transactions and purchases in mobile channels, interactions within these channels, and credit-related features, particularly credit card usage across the national banking system. These results highlight the critical role of mobile engagement and credit in driving customer data-sharing behaviors, providing financial institutions with strategic insights to enhance competitiveness and innovation in the Open Banking environment.

## 1 Introduction

Open Banking (OB) originates from the idea that customers own their data and can share it with other financial institutions to access better services and products [1], [2]. Its implementation started in the EU with the Payment Services Directive (PSD2) 2015. It spread to various countries, including Canada, South Korea, India, Hong Kong, Japan, Singapore, Australia, New Zealand, and Brazil [3]. However, this shift poses risks to financial institutions that do not manage data flow proactively, potentially losing their market share [4].

OB increases competitiveness by enhancing the consumer experience, reducing prices, and improving services [5]. It disrupts traditional market dynamics, where major banks dominate, by lowering switching costs and fostering a data-driven financial services ecosystem [6]. Standardized data exchange and access to transactional information facilitate new market entrants, making competition more viable [4].

Financial institutions can gain a competitive edge by effectively managing data inflow and outflow [7]. Identifying and encouraging customers to share data from other institutions can increase market share, improve services, and provide insights into market dynamics [8]. Conversely, retaining customers likely to share data with competitors is crucial to prevent loss of market share and protect proprietary services.

Despite the benefits, customers are often reluctant to share their financial data due to concerns about fraud, data protection, and cyberattacks [9], [10]. While perceived risk negatively impacts OB adoption, benefits and ease of use can motivate data sharing [11]. Identifying customers willing to share data on a large scale remains challenging. Surveys can be costly and have low response rates, making comprehensive evaluation difficult [12].

Machine learning techniques can predict OB sharing and interpret influential factors [13]. Data mining involves data preparation, model training, and interpretability [14]. Machine learning applications in banking, such as predicting customer churn and credit risk, demonstrate substantial improvements over traditional models [15]. Model interpretability is crucial, and methods like SHAP enhance understanding of decision-making processes [16].

This paper proposes and validates a framework for predicting customers' propensity to share data through OB. The framework uses data from the National Financial System and involves three phases: (i) data preprocessing with ADASYN and NEARMISS algorithms to address class imbalance; (ii) predictive modeling with XGBoost and Bayesian Optimization; and (iii) feature interpretation using SHAP to understand feature contributions. Integrating SHAP values into CART models provides nuanced insights into customer decisions, enhancing transparency and fairness. This approach captures complex data patterns and offers actionable insights into customer behavior in OB, promoting more customer-centric banking practices.

Sections 2 and 3 cover the literature review and framework description, respectively. Section 4 describes the empirical context and dataset, followed by a discussion of results in Section 5 and conclusions in Section 6.

## 2 Literature Review

This section presents the foundational concepts of OB and the techniques utilized in this study. A thorough review of the existing literature highlights the integration of OB within the financial sector and explores various predictive approaches employed.

### 2.1 Open Banking

Financial institutions' data includes balances, daily transactions, financing, loans, consortiums, insurance, credit card bills, investments, and sociodemographic information [17], [18], [19]. However, each organization has its format for structuring and storing this data, without the commitment or interest in following standards that could facilitate sharing it with other financial institutions. In this context, OB establishes a protocol for participating institutions to enable customer data sharing easily, consistently, and securely. This exchange of information occurs through an Application Programming Interface (API), which is defined and managed by a regulatory and supervisory body [20], [21]. The API includes norms, standards, and security regulations for the exchange of information between financial institutions based on customer consent [17].

The sharing process begins when the customer agrees to the transmission of his/her financial transaction data and socio-demographic data from a focal financial institution to its competitors (outflow) or from a competitor to a focal financial institution (inflow) [20]. As a result, the financial institution receiving the data has more information to analyze customers' risk, as well as the potential to better serve those customers, enabling more assertive and attractive product and service offers [22]. Similarly, this data can be analyzed to compose information about market behavior and guide the development of innovative products and services [23], [24]. According to [25], OB data can be applied in various activities such as financial management and personalized marketing. However, in the literature, studies on predictive models using OB are scarce.

Some of those studies used data provided by OB to improve Credit Score calculations. In [13], [22], [26], OB contributed to a more accurate Credit Score estimation, especially for new customers. For that matter, data received from other institutions was combined with techniques like Deep Learning, XGBoost, text mining, and SHAP for model interpretation.

However, there were no observed studies that utilized the predictive potential of data modeling to anticipate customers' decisions to share their data via OB (inflow or outflow). The benefits of receiving data on customers' relationships with competitors highlight the importance of a focal financial institution proactively managing the incentive for data inflow from its customers. Conversely, given the risks of losing market competitiveness when many of its customers promote data outflow to competitors [4], companies must make efforts to reduce or even interrupt this behavior. In this new type of data-driven financial services business ecosystem [6] comprised of millions of customers, the proposition of methods relying on machine learning tools emerges as an important ally in promoting proactive management of customer data-sharing behavior via OB.

### 2.2 Data Preprocessing Methods

In addition to reaching a model yielding high predictive performance, inflow and outflow data also impose the challenge of unbalanced data, given that customers who consent to data sharing end up being a rare class. Unbalanced data compromises the modeling process and, consequently, the accuracy of predictions, as they can lead to biased or overfitted models that do not generalize well to unseen data [27]. To address that issue, data preprocessing is required. Data preprocessing aims to transform raw data into useful information for training a model [28]. [29] claim that data preprocessing comprises the selection, cleaning, and transformation of data. Furthermore, according to [30], this encompasses more than 80% of the modeling effort. In our research, Feature Engineering (FE), Imbalanced Dataset Treatment Oversampling (IDT-over), and Imbalanced Dataset Treatment Undersampling (IDT-under) methods were adopted for preprocessing.

#### 2.2.1 Feature Engineering (FE)

Transforming raw data into a format that enhances model performance is a critical step in predictive modeling [31]. For [32], machine learning models exhibit unique responses to engineered features, significantly influencing their performance. In the study by [33], the pivotal role of FE in enhancing the prediction process is highlighted, particularly in the context of churn prediction within financial institutions. This research underscores FE's significance in refining data preparation and modeling techniques, thus contributing to developing more accurate and reliable predictive models for customer retention. Numerous techniques for generating new features from raw data incorporate the development of frequency and recency features from longitudinal data, guided by the RFM (Recency, Frequency, Monetary value) framework [34]. This allows the expression of customers' behavior related to consumption, enabling a better understanding of how customers interact with products or services over time. Such approaches enhance the granularity of predictive models, particularly in contexts where customer engagement is observed [35].

### 2.2.2 IDT-over

In binary classification modeling, it's common for one of the classes to present substantially fewer observations than the other class (minority and majority classes, respectively). The minority class tends to offer insufficient information for model creation [36]. In this context, oversampling algorithms aim to circumvent this issue by generating synthetic instances that reinforce the boundary between the classes [37]. One such algorithm is the ADASYN, which uses k-nearest neighbor (KNN) to add artificial instances in regions deemed more difficult for class discrimination [38]. ADASYN identifies a weighted distribution of the rare class considering its learning difficulty by using instances of the rare class positioned closely to the majority class. By doing that, it generates other cases that reinforce points on the boundary between the two classes.

### 2.2.3 IDT-under

IDT-under methods remove instances from the majority class to reduce bias caused by a large number of instances in that class [39]. There are various techniques for this process, ranging from simple approaches like random instance removal (RU) to more refined algorithms. However, techniques like RU do not ensure the retention of instances that maximize pattern identification, failing to select significant instances [40]. In this regard, the NEARMISS algorithm aims to remove instances that represent noise or redundancies from the majority class, retaining the most relevant instances for the learning process [41].

## 2.3 XGBoost with Bayesian Hyperparameter Optimization

XGBoost is a machine-learning method that relies on decision trees [42]. It combines multiple weak decision trees to create a robust model employing the Classification and Regression Trees (CART) methodology [43] as its foundation. CART is a technique that facilitates the construction of binary decision trees for classification and regression tasks, optimizing the homogeneity of terminal nodes through successive splits. XGBoost enhances the traditional CART concept by applying an iterative adjustment process: each new tree is sequentially added to correct errors from the previous model. This achievement is accomplished by minimizing a specific loss function, such as the logistic loss function or quadratic loss function, using a gradient optimization approach. XGBoost hyperparameters can be tuned through Bayesian Hyperparameters Optimization [44], which uses a probabilistic model to define hyperparameter values that maximize a chosen performance metric (e.g., PR-AUC). The process is iterative, meaning the algorithm learns to achieve optimal or near-optimal hyperparameter values in each iteration [45], [46].

## 2.4 Explanatory Model Analysis (EMA)

EMA seeks to enhance the interpretability of machine learning models through both model-specific and model-agnostic methods, pivotal for understanding model behavior across various scenarios [47]. Model-specific methods are designed for particular architectures, providing in-depth insights into how these models function. In contrast, model-agnostic methods like LIME (Local Interpretable Model-agnostic Explanations) and SHAP (SHapley Additive exPlanations) are flexible enough to be applied across different models, irrespective of their internal mechanics. These approaches, whether example-based, focusing on individual data points to explain predictions, or feature-based, assessing the impact of each feature as utilized by SHAP, facilitate a nuanced understanding of machine learning. Such exploration not only aids in demystifying complex models but also supports informed decision-making by elucidating the origins of predictions [47]. Building on this foundation, EMA employs established methodologies such as SHAP and LIME to clarify how input features impact model predictions, addressing the imperative to decode complex model behaviors [48], [49]. The implementation of these methodologies significantly enhances transparency and trust in machine learning models, effectively transforming them from "black boxes" into tools that are both comprehensible and reliable [50]. This process not only helps in validating predictions by providing explicit explanations of automated decisions but also assists in identifying and amending biases and errors, thereby ensuring fair and accurate outcomes [51]. Moreover, the insights garnered from EMA can inform model improvement and business strategies while promoting effective communication between data scientists and stakeholders [52].

SHAP is a method for interpreting machine learning models rooted in cooperative game theory [53]. This approach has gained recognition due to its ability to provide clear and consistent interpretations of predictions generated by complex models. Cooperative game theory provides the conceptual framework for calculating the Shapley value of each feature, reflecting its marginal contribution to the final prediction [13]. SHAP rationale relies on the notion that each independent feature uniquely and measurably contributes to a model's prediction. SHAP evaluates the importance of each feature, considering all possible combinations of features and calculating the weighted average of marginal contributions, resulting in a global explanation and individualized interpretation for each instance [48]. It quantifies the impact of including or excluding a feature by averaging its effect across all possible feature combinations, leading to a comprehensive understanding of its contribution to the prediction outcome. Unlike the Relative Gain metric from XGBoost, which may disproportionately highlight the importance of top features due to its aggregation across all trees in the model [54], SHAP offers a balanced measure of feature importance. When SHAP is employed alongside a logistic XGBoost model, the outcome is an additive model for each instance, as outlined in Equations (1) and (2) [54].

$$logit\ (p)_i = \beta_0 + \sum_{j=1}^{n} \varphi_{ij} \qquad (1)$$

Where $i$ is the instance, $j$ is the feature, $n$ is the number of features, $\varphi$ is a SHAP Value, and the result is $logit\ (p)$. $\beta_0$ is the base value, the mean prediction of the model across a reference dataset, serving as the baseline for calculating feature contributions.

$$p_i = \frac{1}{1 + e^{-logit(p)_i}} \qquad (2)$$

Where $i$ is the instance, $e$ is the Euler's number, and $p$ represents the probability of the event is true.

# 3 Proposed Method

This section presents a framework for predicting and analyzing data sharing in the OB context, aiming to predict customer data sharing propensity and identify influential features. The framework, applicable to data inflow and outflow models, consists of three phases: data preprocessing, model training and testing, and Explanatory Model Analysis (EMA) (see Fig. 1). This comprehensive approach spans initial data handling to the detailed interpretation of predictive models, addressing data flow within financial institutions and providing a holistic view of customer behavior in OB.

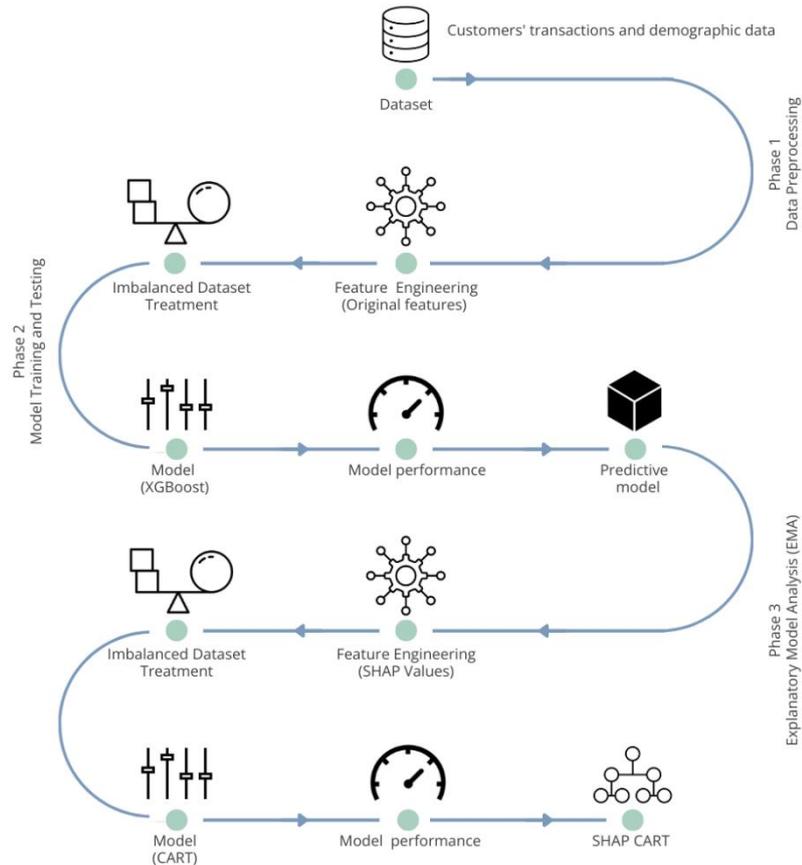

**Fig. 1.** Proposed method.

## 3.1 Phase 1 - Data Preprocessing

This phase involves three preprocessing steps.

### 3.1.1 Step 1 - FE Preprocessing

The original dataset includes customer transaction histories and socio-demographic details. Data transformation into features adhering to the RFM (Recency, Frequency, Monetary value) concept allows for a detailed analysis of customer engagement. This method enhances predictive accuracy by incorporating behavioral indicators and credit standings.

### 3.1.2 Step 2 - IDT-over and IDT-under Preprocessing

To balance the dataset, the ADASYN algorithm generates synthetic instances of customers who have shared data (rare class). The Kolmogorov-Smirnov test ensures no significant distribution changes. The NEARMISS algorithm then removes redundant majority class instances, verified again using the Kolmogorov-Smirnov test.

## 3.2 Phase 2 - Model Training and Testing

XGBoost models are constructed to predict customer data-sharing propensity. Bayesian Hyperparameter Optimization, using the "tune" package in R, optimizes parameters such as tree number, node splitting, tree depth, and adaptation rate. Model evaluation employs metrics like accuracy, specificity, PR-AUC, and recall using 100-fold cross-validation.

## 3.3 Phase 3 - Explanatory Model Analysis (EMA)

### 3.3.1 Step 1 - SHAP Values

SHAP (Shapley Additive Explanations) values elucidate each feature's prediction impact. After retraining XGBoost models, SHAP values are calculated across the entire dataset using R packages "SHAPforxgboost" and "shapviz".

### 3.3.2 Step 2 - SHAP Values CART

Classification and Regression Trees (CART) use SHAP values as primary predictive features, providing an intuitive understanding of feature contributions. CART models, validated using 10-fold cross-validation, enhance model interpretability and identify misclassification patterns to improve model performance.

The proposed framework offers a robust methodology for predicting and analyzing data sharing in OB, fostering customer-centric banking practices through detailed feature interpretation and model transparency.

# 4 Data and Empirical Context

In Brazil, the OB initiative, termed Open Finance by the Central Bank of Brazil [55], includes a wider range of financial entities beyond traditional banks. This broader scope enhances the spectrum of financial services and data-sharing capabilities. This research focuses on customers who have completed all stages of data sharing within the Open Finance framework, designated as 'true inflow' or 'true outflow' based on their data-sharing activities. Data was collected from a major Brazilian bank offering various financial services, including private pensions, credit cards, loans, real estate and vehicle financing, investments, and insurance. The study examines longitudinal transaction data over 36 months (January 2020 to December 2022), covering approximately 203 million transactions from 3,244,698 inflow and 3,254,168 outflow customers. Differences in customer numbers are due to the exclusion of individuals who had not completed data sharing. Additionally, data from the Brazilian National Financial System provides insights into customer market debt, accessible to all financial entities recognized by the Central Bank of Brazil [56]. A significant finding is the class imbalance: only 0.093% of customers imported data (inflow), and 0.38% exported data (outflow) through Open Finance. This structured data analysis supports the study's examination of Open Finance's impact on customer data-sharing behavior.

## 5   Results and Discussion

### 5.1   Predictive performance

The proposed framework was used to develop two models to understand customer decisions regarding data sharing in OB. The Outflow Model predicts the probability of a customer sharing their data for the first time, enabling data transmission from the focal bank to another financial institution. The Inflow Model predicts the probability of a customer importing data from another financial institution to the focal bank.

We evaluated each model's predictive performance using averages and standard deviations of PR-AUC, recall, specificity, and accuracy (see Table 1). To ensure reliability, we employed 100-fold cross-validation, assessing performance across various scenarios for robust generalizability estimates.

Table 1. Predictive performance metrics

| Model | PR-AUC | Recall | Specificity | Accuracy |
| --- | --- | --- | --- | --- |
| Outflow model | .91537 (.00817) | .88183(.02745) | .80929 (.00270) | .80957 (.00269) |
| Inflow Model | .91398 (.01895) | .86250 (.05676) | .81789 (.00403) | .81793 (.00403) |

The outflow model, designed to predict a customer's initial decision to share data with an external financial institution, showed strong predictive ability. It achieved a mean PR-AUC of 0.91537, reflecting balanced precision and recall. The average recall was 0.88183, indicating effective identification of actual positives. Specificity and accuracy both averaged above 0.80, demonstrating accurate differentiation of customer decisions. The low standard deviation across metrics indicates consistent performance.

The Inflow Model, predicting the probability of customers importing data into the focal bank, also demonstrated notable performance. The mean PR-AUC was 0.91398, indicating effectiveness in identifying customers likely to share their data. Although the mean recall and specificity were slightly lower, at 0.86250 and 0.81789 respectively, they remain high, affirming practical relevance. An accuracy of 0.81793 supports its predictive precision, and the higher standard deviation still points to acceptable consistency.

### 5.2   Explanatory Model Analysis (EMA)

#### 5.2.1   SHAP Feature Importance Ranking

Feature importance rankings are crucial for interpreting predictive models, providing insights into the factors driving model decisions, and guiding strategic business decisions. In the outflow model, SHAP values show a more balanced distribution, with 80% of the importance spanning 33 features. "Number of interactions in mobile channels" is the leading feature with 16.72% importance, followed by "Number of transactions and purchases in mobile channels" at 8.07%, and "Number of interactions in digital channels" at 6.55%. This balanced attribution of feature contribution illustrates a more integrative assessment within the SHAP framework.

For the inflow model, "Number of transactions and purchases in mobile channels" holds 11.14% SHAP importance, suggesting a balanced ranking among top features. "Number of interactions in mobile channels" follows closely with 10.68% importance, indicating a similar significance between the top two features.

These observations from both models highlight the balanced nature of SHAP values in representing feature importance. While some methods may disproportionately emphasize single features, SHAP provides a holistic assessment of feature relevance. This comprehensive view is essential for targeted interventions and a thorough understanding of the predictive dynamics within models.

### 5.2.2  SHAP Values CART

Within the outflow model, "Digital maturity - no digital activity" is the predominant feature in the CART decision tree. SHAP values below 0.069 indicate non-outflow occurrences, suggesting that customers who do not use the bank's digital resources are less likely to exhibit outflow. Conversely, customers with any digital activity have a higher chance of data outflow. Additional significant features include "Overall credit value" and "Credit value taken in the whole national banking system - credit cards".  Both are characterized by negative SHAP values regarding outflow. Customers with minimal credit availability and unmet credit needs from the focal bank tend to seek additional credit elsewhere, initiating outflow. This increases their susceptibility to potential delinquency, impacting their financial stability within the focal bank and other financial institutions from which they may seek additional credit.

In the inflow model, the CART decision tree analysis reveals a greater number of branches compared to the outflow model. The "Number of transactions and purchases in mobile channels" and the "Number of interactions in mobile channels" are key determinants. There is a positive correlation between the frequency of mobile channel use and SHAP values contributing to inflow. Since OB relies on interaction with mobile apps, customers adept at using mobile services show a greater propensity for inflow. Credit-related features, such as overdue credit tied to National System Credit Cards, indicate customers' pursuit of additional credit resources. Additionally, customers with master's degrees tend to show less necessity for credit. Despite their proficiency in using OB, they are not actively seeking additional credit.

## 6   Final Considerations

This study presents a framework for predicting and explaining customer data sharing in OB, enhancing understanding of customer behavior in financial markets. Using SHAP on the XGBoost algorithm, our approach shows high predictive performance. This is crucial for financial institutions to maximize OB's potential. The framework identifies customers likely to share data and provides insights into their decisions, helping managers develop strategies to enhance customer loyalty and optimize services.

Applying our framework in a major Brazilian retail bank highlights its effectiveness, with significant implications for research and industry practice. Achieving 91.53% PR-AUC for data outflow and 91.39% PR-AUC for inflow propensity, this study sets a new benchmark in banking predictive analytics. Using SHAP for model interpretation also enhances decision-making with a detailed understanding of customer behavior.

Future research should focus on longitudinal analyses of customer data-sharing behavior and the framework's scalability across different markets. As OB evolves, understanding customer data sharing is critical for innovation and financial inclusivity globally.

## References

[1]   Q. Tang, G. Xia, X. Zhang, e F. Long, "A Customer Churn Prediction Model Based on XGBoost and MLP", em *2020 International Conference on Computer Engineering and Application (ICCEA)*, mar. 2020, p. 608–612. doi: https://doi.org/10.1109/ICCEA50009.2020.00133.

[2] D. Zetzsche, D. Arner, R. Buckley, e R. H. Weber, "The Evolution and Future of Data-Driven Finance in the EU", *Common Market L. Rev.*, vol. 57, p. 331, 2020.

[3] M. Kassab e P. A. Laplante, "Open Banking: What It Is, Where It's at, and Where It's Going", *Computer*, vol. 55, nº 1, p. 53–63, jan. 2022, doi: 10.1109/MC.2021.3108402.

[4] C. Frei, "Open Banking: Opportunities and Risks", em *The Fintech Disruption: How Financial Innovation Is Transforming the Banking Industry*, T. Walker, E. Nikbakht, e M. Kooli, Orgs., em Palgrave Studies in Financial Services Technology. , Cham: Springer International Publishing, 2023, p. 167–189. doi: 10.1007/978-3-031-23069-1_7.

[5] D. W. Arner, R. P. Buckley, e D. A. Zetzsche, "Open Banking, Open Data, and Open Finance: Lessons from the European Union", em *Open Banking*, L. Jeng, Org., Oxford University Press, 2022, p. 0. doi: 10.1093/oso/9780197582879.003.0009.

[6] D. W. Arner, R. P. Buckley, e D. A. Zetzsche, "Open Banking, Open Data and Open Finance: Lessons from the European Union", 11 de novembro de 2021, *Rochester, NY*: 3961235. Acesso em: 9 de julho de 2023. [Online]. Disponível em: https://papers.ssrn.com/abstract=3961235

[7] S. Mansfield-Devine, "Open banking: opportunity and danger", *Computer Fraud & Security*, vol. 2016, nº 10, p. 8–13, out. 2016, doi: 10.1016/S1361-3723(16)30080-X.

[8] Z. He, J. Huang, e J. Zhou, "Open banking: Credit market competition when borrowers own the data", *Journal of Financial Economics*, vol. 147, nº 2, p. 449–474, fev. 2023, doi: 10.1016/j.jfineco.2022.12.003.

[9] O. Borgogno e G. Colangelo, "The data sharing paradox: BigTechs in finance", p. 1–25, 28 de maio de 2020.

[10] G. Littlejohn, G. Boskovich, e R. Prior, "United Kingdom: The Butterfly Effect", em *Open Banking*, L. Jeng, Org., Oxford University Press, 2022, p. 0. doi: 10.1093/oso/9780197582879.003.0010.

[11] R. Chan, I. Troshani, S. Rao Hill, e A. Hoffmann, "Towards an understanding of consumers' FinTech adoption: the case of Open Banking", *International Journal of Bank Marketing*, vol. 40, nº 4, p. 886–917, jan. 2022, doi: 10.1108/IJBM-08-2021-0397.

[12] C. Asif, T. Olanrewaju, H. Sayama, e A. Vijayasrinivasan, "Financial services unchained: The ongoing rise of open banking", Mckinsey & Company. Acesso em: 13 de julho de 2023. [Online]. Disponível em: https://www.mckinsey.com/industries/financial-services/our-insights/financial-services-unchained-the-ongoing-rise-of-open-financial-data

[13] L. O. Hjelkrem e P. E. de Lange, "Explaining Deep Learning Models for Credit Scoring with SHAP: A Case Study Using Open Banking Data", *Journal of Risk and Financial Management*, vol. 16, nº 4, Art. nº 4, abr. 2023, doi: 10.3390/jrfm16040221.

[14] F. Gorunescu, "Introduction to Data Mining", em *Data Mining: Concepts, Models and Techniques*, F. Gorunescu, Org., em Intelligent Systems Reference Library. , Berlin, Heidelberg: Springer, 2011, p. 1–43. doi: 10.1007/978-3-642-19721-5_1.

[15] R. A. de L. Lemos, T. C. Silva, e B. M. Tabak, "Propension to customer churn in a financial institution: a machine learning approach", *Neural Comput & Applic*, mar. 2022, doi: 10.1007/s00521-022-07067-x.

[16] P. Fukas, J. Rebstadt, L. Menzel, e O. Thomas, "Towards Explainable Artificial Intelligence in Financial Fraud Detection: Using Shapley Additive Explanations to Explore Feature Importance", em *Advanced Information Systems Engineering*, X. Franch, G. Poels, F. Gailly, e M. Snoeck, Orgs., Cham: Springer International Publishing, 2022, p. 109–126. doi: 10.1007/978-3-031-07472-1_7.

[17] G. Long, Y. Tan, J. Jiang, e C. Zhang, "Federated Learning for Open Banking", em *Federated Learning: Privacy and Incentive*, Q. Yang, L. Fan, e H. Yu, Orgs., em Lecture Notes in

[18] G. Quatrochi, A. L. G. da Silva, e J. E. Cassiolato, "Banks 4.0 in Brazil: possibilities to ensure fintechs financing role through its market positioning", *Innovation and Development*, vol. 0, nº 0, p. 1–21, jun. 2022, doi: 10.1080/2157930X.2022.2086336.

[19] C.-H. Tsai e K.-J. Peng, *Regulating Open Banking: Comparative Analysis of the EU, the UK and Taiwan*. Taylor & Francis, 2022.

[20] P. Laplante e N. Kshetri, "Open Banking: Definition and Description", *Computer*, vol. 54, nº 10, p. 122–128, out. 2021.

[21] B. Ramdani, B. Rothwell, e E. Boukrami, "Open Banking: The Emergence of New Digital Business Models", *Int. J. Innovation Technol. Management*, vol. 17, nº 05, p. 2050033, ago. 2020, doi: 10.1142/S0219877020500339.

[22] L. O. Hjelkrem, P. E. de Lange, e E. Nesset, "The Value of Open Banking Data for Application Credit Scoring: Case Study of a Norwegian Bank", *Journal of Risk and Financial Management*, vol. 15, nº 12, Art. nº 12, dez. 2022, doi: 10.3390/jrfm15120597.

[23] A. M. Dahdal e B. Zeller, "Open Banking and Open Data: Context, Innovation, and Consumer Protection", *Banking L.J.*, vol. 138, p. 385, 2021.

[24] T. Lynn, P. Rosati, e M. Cummins, "Exploring Open Banking and Banking-as-a-Platform: Opportunities and Risks for Emerging Markets", em *Entrepreneurial Finance in Emerging Markets: Exploring Tools, Techniques, and Innovative Technologies*, D. Klonowski, Org., Cham: Springer International Publishing, 2020, p. 319–334. doi: 10.1007/978-3-030-46220-8_20.

[25] E. E. J. P. Hassany e G. T. Pambekti, "Review on The Application of Open Banking in Sharia Banking: An Swot Analysis", *Jurnal Ekonomi dan Keuangan Syariah*, vol. 5, 2022, [Online]. Disponível em: https://scholar.archive.org/work/4j7hn4htkfay5mxqa3vf27j2he/access/wayback/https://ojs.unimal.ac.id/el-amwal/article/download/6676/pdf

[26] L. O. Hjelkrem, P. Lange, e E. Nesset, "An end-to-end deep learning approach to credit scoring using CNN + XGBoost on transaction data", 21 de julho de 2022, *Rochester, NY*: 4168935. Acesso em: 29 de junho de 2023. [Online]. Disponível em: https://papers.ssrn.com/abstract=4168935

[27] R. Bafna, R. Jain, e R. Malhotra, "A Comparative Study of Classification Techniques and Imbalanced Data Treatment for Prediction of Software Faults", *Research Square*, abr. 2023, doi: https://doi.org/10.21203/rs.3.rs-2809140/v1.

[28] C. Sammut e G. I. Webb, Orgs., "Data Preprocessing", em *Encyclopedia of Machine Learning*, Boston, MA: Springer US, 2010, p. 260–260. doi: 10.1007/978-0-387-30164-8_195.

[29] S. García, J. Luengo, e F. Herrera, *Data Preprocessing in Data Mining*. Springer International Publishing, 2014.

[30] D. Pyle, *Data Preparation for Data Mining (The Morgan Kaufmann Series in Data Management Systems)*, Book&CD-ROM 1st. em The Morgan Kaufmann Series in Data Management Systems. Morgan Kaufmann, 1999.

[31] P. Duboue, *The Art of Feature Engineering: Essentials for Machine Learning*. Cambridge University Press, 2020.

[32] J. Heaton, "An empirical analysis of feature engineering for predictive modeling", em *SoutheastCon 2016*, mar. 2016, p. 1–6. doi: 10.1109/SECON.2016.7506650.

[33] Brito *et al.*, "A framework to improve churn prediction performance in retail banking", *Financial Innovation*, vol. 10, nº 1, p. 17, jan. 2024, doi: 10.1186/s40854-023-00558-3.

[34] K. Singh, D. Arora, e P. Sharma, "Identification of Shoplifting Theft Activity Through Contour Displacement Using OpenCV", em *Computational Methods and Data Engineering*, V. Singh, V. K. Asari, S. Kumar, e R. B. Patel, Orgs., em Advances in Intelligent Systems and Computing. Singapore: Springer, 2021, p. 441–450. doi: 10.1007/978-981-15-6876-3_34.

[35] R. Heldt, C. S. Silveira, e F. B. Luce, "Predicting customer value per product: From RFM to RFM/P", *Journal of Business Research*, vol. 127, p. 444–453, abr. 2021.

[36] G. M. Weiss, "Mining with rarity: a unifying framework", *SIGKDD Explor. Newsl.*, vol. 6, nº 1, p. 7–19, 2004, doi: https://doi.org/10.1145/1007730.1007734.

[37] I. Triguero, J. Derrac, S. Garcia, e F. Herrera, "A Taxonomy and Experimental Study on Prototype Generation for Nearest Neighbor Classification", *IEEE Trans. Syst., Man, Cybern. C*, vol. 42, nº 1, p. 86–100, 2012, doi: https://doi.org/10.1109/TSMCC.2010.2103939.

[38] H. He, Y. Bai, E. A. Garcia, e S. Li, "ADASYN: Adaptive synthetic sampling approach for imbalanced learning", em *2008 IEEE International Joint Conference on Neural Networks (IEEE World Congress on Computational Intelligence)*, jun. 2008, p. 1322–1328. doi: 10.1109/IJCNN.2008.4633969.

[39] A. Fernandez, S. Garcia, F. Herrera, e N. V. Chawla, "SMOTE for Learning from Imbalanced Data: Progress and Challenges, Marking the 15-year Anniversary", *Journal of Artificial Intelligence Research*, vol. 61, p. 863–905, abr. 2018, doi: 10.1613/jair.1.11192.

[40] H. He e Y. Ma, Orgs., *Imbalanced learning: foundations, algorithms, and applications*. Hoboken, New Jersey: John Wiley & Sons, Inc, 2013.

[41] J. Zhang e I. Mani, "KNN Approach to Unbalanced Data Distributions: A Case Study Involving Information Extraction", em *Proceeding of International Conference on Machine Learning*, Washington DC: ICML United States, 2003. [Online]. Disponível em: https://www.site.uottawa.ca/~nat/Workshop2003/jzhang.pdf

[42] T. Chen e C. Guestrin, "XGBoost: A Scalable Tree Boosting System", em *Proceedings of the 22nd ACM SIGKDD International Conference on Knowledge Discovery and Data Mining*, em KDD '16. New York, NY, USA: Association for Computing Machinery, Agosto 2016, p. 785–794.

[43] L. Breiman, J. Friedman, C. J. Stone, e R. A. Olshen, *Classification and Regression Trees*. Routledge, 2017.

[44] A. H. Victoria e G. Maragatham, "Automatic tuning of hyperparameters using Bayesian optimization", *Evolving Systems*, vol. 12, nº 1, p. 217–223, mar. 2021, doi: https://doi.org/10.1007/s12530-020-09345-2.

[45] J. Snoek, H. Larochelle, e R. P. Adams, "Practical bayesian optimization of machine learning algorithms", em *Advances in neural information processing systems 25 (NIPS 2012)*, NeurIPS Proceedings, 2012.

[46] Y. Xia, C. Liu, Y. Li, e N. Liu, "A boosted decision tree approach using Bayesian hyper-parameter optimization for credit scoring", *Expert Systems with Applications*, vol. 78, p. 225–241, jul. 2017, doi: 10.1016/j.eswa.2017.02.017.

[47] K. W. De Bock *et al.*, "Explainable AI for Operational Research: A defining framework, methods, applications, and a research agenda", *European Journal of Operational Research*, set. 2023, doi: 10.1016/j.ejor.2023.09.026.

[48] S. M. Lundberg e S.-I. Lee, "A unified approach to interpreting model predictions", em *Proceedings of the 31st International Conference on Neural Information Processing Systems*, em NIPS'17. Red Hook, NY, USA: Curran Associates Inc., Dezembro 2017, p. 4768–4777.

[49] M. Ribeiro, S. Singh, e C. Guestrin, "'Why Should I Trust You?': Explaining the Predictions of Any Classifier", em *Proceedings of the 2016 Conference of the North American Chapter of the Association for Computational Linguistics: Demonstrations*, J. DeNero, M. Finlayson, e S. Reddy, Orgs., San Diego, California: Association for Computational Linguistics, jun. 2016, p. 97–101. doi: 10.18653/v1/N16-3020.

[50] A. Nandi e A. K. Pal, *Interpreting Machine Learning Models: Learn Model Interpretability and Explainability Methods*. Apress, 2021.

[51] P. Biecek e T. Burzykowski, *Explanatory Model Analysis: Explore, Explain, and Examine Predictive Models*. CRC Press, 2022.

[52] U. Bhatt *et al.*, "Explainable machine learning in deployment", em *Proceedings of the 2020 Conference on Fairness, Accountability, and Transparency*, em FAT* '20. New York, NY, USA: Association for Computing Machinery, jan. 2020, p. 648–657. doi: 10.1145/3351095.3375624.

[53] L. S. Shapley, "A Value for n-Person Games", em *Contributions to the Theory of Games (AM-28)*, vol. II, Princeton University Press, 1953. [Online]. Disponível em: https://doi.org/10.1515/9781400881970-018

[54] T. Chen *et al.*, "xgboost: Extreme Gradient Boosting", *CRAN*, vol. R package version 1.7.7.1, 2024, doi: https://CRAN.R-project.org/package=xgboost.

[55] OF BACEN, "Open Finance Brasil", Open Finance Brasil. Acesso em: 9 de fevereiro de 2024. [Online]. Disponível em: https://www.bcb.gov.br/estabilidadefinanceira/sfn

[56] B. SFN, "Sistema Financeiro Nacional (SFN)", Sistema Financeiro Nacional (SFN). Acesso em: 9 de fevereiro de 2024. [Online]. Disponível em: https://www.bcb.gov.br/estabilidadefinanceira/sfn